\documentclass[11pt,onecolumn,amssymb,nofootinbib]{revtex4}
\usepackage{amsmath, amsthm, amscd, amssymb}
\usepackage{graphicx, braket}
\usepackage{bm}
\usepackage{bbm}

\begin{document}

\title{\bf The quantum theory of measurement within dynamical
reduction models}
\author{Angelo Bassi}
\email{bassi@ts.infn.it,  bassi@mathematik.uni-muenchen.de}
\address{Dipartimento di Fisica Teorica,
Universit\`a di Trieste, Strada Costiera 11, 34014 Trieste, Italy.
\\  Mathematisches Institut der L.M.U., Theresienstr. 39, 80333
M\"unchen, Germany. \\ Istituto Nazionale di Fisica Nucleare,
Sezione di Trieste, Strada Costiera 11, 34014 Trieste, Italy.}
\author{Davide~G.~M. Salvetti}
\email{salvetti@ts.infn.it} \affiliation{Dipartimento di Fisica
Teorica dell'Universit\`a Trieste, Strada Costiera 11, 34014
Trieste, Italy. \\ Istituto Nazionale di Fisica Nucleare, Sezione di
Trieste, Strada Costiera 11, 34014 Trieste, Italy.}
\begin{abstract}
We analyze in mathematical detail, within the framework of the QMUPL
model of spontaneous wave function collapse, the von Neumann
measurement scheme for the measurement of a $1/2$ spin particle. We
prove that, according to the equation of the model: i) throughout
the whole measurement process, the pointer of the measuring device
is always perfectly well localized in space; ii) the probabilities
for the possible outcomes are distributed in agreement with the Born
probability rule; iii) at the end of the measurement the state of
the microscopic system has collapsed to the eigenstate corresponding
to the measured eigenvalue. This analysis shows rigorously how
dynamical reduction models provide a consistent solution to the
measurement problem of quantum mechanics.
\end{abstract}
\maketitle

\section{Introduction}
\label{sec:introduction}

In standard textbooks on Quantum Mechanics, e.g. in~\cite{mes}, one
can find the following axioms defining the quantum theory:\\

\noindent {\textsc{Axiom 1: states.}  A Hilbert ${\mathcal H}$ space
is associated to each physical system and the state of the system is
represented by a vector $|\psi\rangle$ in ${\mathcal H}$. (In the
following, we will always assume vectors to be normalized.)\\

\noindent {\textsc{Axiom 2: observables.} To any observable quantity
of the system is associated a self-adjoint operator in ${\mathcal
H}$. The only possible outcomes of a measurement of an observable
are the eingenvalues of the associated operator.\\

\noindent {\textsc{Axiom 3: Schr\"odinger equation.} Given
$|\psi_{0}\rangle$ the state of the system at an initial time $t_{0}
= 0$, its state at any subsequent time $t$ is represented by
$|\psi_{t}\rangle$, which is the solution of the Schr\"odinger
equation:
\begin{equation} \label{eq:sch}
i\hbar\,\frac{d}{dt}\, |\psi_{t}\rangle \; = \; H\,
|\psi_{t}\rangle,
\end{equation}
for the given initial condition; the self-adjoint operator $H$ is
the Hamiltonian operator associated to the system.\\

\noindent {\textsc{Axiom 4: Born rule.} Let $|\psi\rangle$ be the
vector describing the state of the system at a given time; then the
probability that the outcome of a measurement of an observable
${\mathcal A}$ at that time is one of the values $a_{n}$ belonging
to the spectrum of $A$, is given by the Born probability rule:
\begin{equation}
{\mathbb P}[a_{n}] \quad = \quad \langle \psi| P_{n} | \psi\rangle,
\end{equation}
where $P_{n}$ is the projection operator associated to the
eigenmanifold of the operator $A$ corresponding to the eigenvalue
$a_{n}$.\\

\noindent {\textsc{Axiom 5: wave-packet reduction.} At the end of a
measurement process the state of the system changes according to the
rule:
\begin{equation}
|\psi\rangle \quad \xrightarrow[\makebox{\tiny after measurement}]{}
\quad \frac{P_{n} |\psi\rangle}{\| P_{n} |\psi\rangle \| },
\end{equation}
where $P_{n}$ is the projection operator associated to the outcome
$a_{n}$ of the measurement.

As well known, the last axiom gives rise to the measurement problem
in Quantum Mechanics, because of which the theory, as it stands,
cannot be considered a consistent description of physical phenomena.
Many tentative solutions have been suggested, among which dynamical
reduction models are one of the few promising proposals; their
general structure has been already fully described in the past
literature~\cite{rev1,rev2}; here we limit ourselves to list the
axioms defining them (at the non relativistic level):\\

\noindent {\textsc{Axiom A: states.} A Hilbert ${\mathcal H}$ space
is associated to each physical system and the state of the system is
represented by a (normalized) vector $|\psi\rangle$ in ${\mathcal
H}$.\\

\noindent {\textsc{Axiom B: evolution (continuous version).}  Given
the system initially in a state described by the vector
$|\psi_{0}\rangle$, its state at any subsequent time $t$ is
represented by $|\psi_{t}\rangle$, which solves the following
stochastically modified Schr\"odinger equation:
\begin{equation} \label{eq:gen}
d\, |\psi_{t}\rangle \; = \; \left[ - \frac{i}{\hbar} H dt \; + \;
\sqrt{\lambda} (A - \langle A \rangle_{t})\, dW_{t} \; - \;
\frac{\lambda}{2} (A - \langle A \rangle_{t})^2 dt
\right]|\psi_{t}\rangle,
\end{equation}
where $W_{t}$ is a standard Wiener process defined on a probability
space $(\Omega, {\mathcal F}, {\mathbb P})$, while $\langle A
\rangle_{t} \equiv \langle \psi_{t} | A | \psi_{t}\rangle$ is the
quantum average value of the operator $A$, which is a suitably
chosen (in this case, self-adjoint) operator; $\lambda$ is a
positive constant controlling the strength of the
collapse\footnote{Eq.~\eqref{eq:gen} can be generalized in different
directions~\cite{rev1,rev2}; however, its general structure has to
be preserved in order for the model to provide a solution to the
measurement problem.}.\\

\noindent {\textsc{Axiom C: ontology.} Let $\psi({\bf x}_{1}, {\bf
x}_{2}, \ldots {\bf x}_{N}) \equiv \langle {\bf x}_{1}, {\bf x}_{2},
\ldots {\bf x}_{N}| \psi\rangle$ the wave function for a system of
$N$ particles (which for simplicity we take to be scalar) in
configuration space. Then
\begin{equation}
\mu_{t}^{(n)}({\bf x}_{n}) \; \equiv \; m_{n} \int d^3 x_{1} \ldots
d^3 x_{n-1} d^2 x_{n+1} \ldots d^3 x_{N} \, | \psi({\bf x}_{1}, {\bf
x}_{2}, \ldots {\bf x}_{N}) |^2
\end{equation}
represents the {\it density of mass}\footnote{In the subsequent
sections, for simplicity's sake, we will not make reference to the
mass density function anymore, but we will only keep track of the
evolution of the wave function; however it should be clear that, in
order to be fully rigorous, all statements about the properties of
physical systems should be phrased in terms of their mass-density
distribution, not in terms of the wave function.} of the $n$-th
particle of the system, to which a total mass $m_{n}$ is
associated~\cite{ggb,pio}.

Axiom A is equal to 1, while B replaces 3. Indeed, B embodies 3,
meaning with this that a sensible choice for $\lambda$ and $A$ can
be made such that Eq.~\eqref{eq:gen} practically reduces to
Eq.~\eqref{eq:sch} when a microscopic quantum system is taken into
account: see~\cite{ad1} for a recent and exhaustive review of the
subject.

The remarkable property of collapse models is that also the other
axioms of Quantum Mechanics derive from B (and C, of course): the
aim of this paper is to show how axioms 4 and 5 derive from axiom B,
while in a future paper we will discuss how axiom 2 also derives
from B. To be more precise, following the previous work of
Ref.~\cite{gmis}, we here analyze in mathematical detail, within the
framework of a specific dynamical reduction model, a von Neumann
type of measurement scheme, in which a microscopic system interacts
with a macroscopic apparatus devised in such a way to measure one or
more properties of the micro-system. We will show, giving also
precise estimates, that:
\begin{enumerate}
\item\label{item:1} whichever the initial state of the microscopic system,
throughout the entire measurement process the apparatus has a
definite position in space, its wave function being always extremely
well localized;
\item\label{item:2} the only possible outcomes correspond to those given
by standard quantum mechanics, with probability almost equal to~$1$;
\item\label{item:3} the probability of getting a certain outcome is given
by the {\it Born probability rule} within an exceedingly high degree
of approximation;
\item\label{item:4} after the measurement, the state vector of the
microscopic system collapses to a state which practically coincides
with the eigenstate of the measured observable, corresponding to the
eigenvalue which has been observed.
\end{enumerate}
Needless to say, these properties were already known since very long
time, and indeed they represent the very motivation behind the
original GRW model~\cite{grw} and its subsequent generalizations,
and the reason for its success; our goal here is to derive them in a
rigorous mathematical way from the equations of a specific model of
wave function collapse.

The paper is organized as follows.  In
section~\ref{sec:measurement-model} we will introduce the
measurement model our analysis is based upon, and we will discuss
its physical features.  In section~\ref{sec:eigenstate-measurement}
we will study the special case in which the microscopic system has
been prepared in an eigenstate of the operator associated to the
observable the model is devised to measure, while in
section~\ref{sec:superposition-measurement} we will analyze in full
detail the case of an arbitrary initial state. In the concluding
section~\ref{sec:conclusions} we will summarize the features of our
model and draw our final conclusions.

\section{The measurement model}
\label{sec:measurement-model} We begin our discussion by presenting
the measurement model we will use in the following sections: the
setup consists of a microscopic system~${\mathcal S}$ interacting
with a macroscopic system~${\mathcal A}$ which acts like a measuring
apparatus; both systems are described in quantum mechanical terms.
Here below we give the details.

\subsection{The microscopic system}
\label{sec:microscopic-system} We consider a single measurement
process, in which the experimenter is able to distinguish among a
{\it finite} set of outcomes.  Accordingly, we assume that the
microscopic system~${\mathcal S}$ can be described, for what
concerns the measurement process, by a finite-dimensional complex
Hilbert space. For the sake of simplicity, and without loss of
generality, we can consider the simplest case:
$\mathcal{H}_{{\mathcal S}} = \mathbb{C}^{2}$, because the
generalization of what follows to~$\mathbb{C}^{n}$ is quite
straightforward.  Since the most general self-adjoint operator~$O$
acting on~$\mathbb{C}^{2}$ can be written as
\begin{equation}
  \label{eq:1}
  O = o_{+} | + \rangle\langle +| + o_{-} | - \rangle\langle -|,
\end{equation}
where~$| + \rangle$ and~$| - \rangle$ are the eigenstates of $O$,
while $o_{+}$ and $o_{-}$ are its two real eigenvalues, for
definiteness and with no loss of generality, in what follows we will
take $o_{\pm} = \pm \hbar / 2$ and $O$ to be the $z$-component of
the spin, $S_{z}$, of a $1/2$ spin particle.

\subsection{The measuring apparatus}
\label{sec:measuring-apparatus} We take the following model for the
measuring apparatus~${\mathcal A}$, which is general enough to
describe all interesting physical situations: we assume that the
apparatus consists of a fixed part plus a pointer moving along a
graduate scale, in such a way that different positions of the
pointer along the scale correspond to different possible outcomes of
the measurement.  To simplify the analysis, we study the evolution
of the center of mass of the pointer only, and disregard all other
macroscopic and microscopic degrees of freedom; accordingly, the
pointer will be treated like a macroscopic quantum particle of
mass~$m$ moving in one dimension only, whose state space is
described by the Hilbert space $\mathcal{H}_{{\mathcal A}} =
\mathrm{L}^{2} (\mathbb{R})$.

\subsection{The Dynamics}
\label{sec:dynamics} We assume that the pointer of~${\mathcal A}$
undergoes a spontaneous collapse mechanism according to the Quantum
Mechanics with Universal Position Localization (QMUPL) model first
introduced in~\cite{di1} and subsequently analyzed in~\cite{ab1}
(see also references therein), while the microscopic
system~${\mathcal S }$ evolves according to the standard
Schr\"odinger equation, since, as typical of dynamical reduction
models, the stochastic collapse terms have very little effects on
microscopic quantum systems. Accordingly, we take for the evolution
equation of the composite ${\mathcal S} + {\mathcal A}$ system the
following stochastic differential equation\footnote{See~\cite{th1}
and~\cite{th2} for theorems on the existence and uniqueness of
solution for this type of equation.} (SDE) defined in the Hilbert
space $\mathcal{H} = \mathcal{H}_{{\mathcal S}} \otimes
\mathcal{H}_{{\mathcal A}} := \mathbb{C}^{2} \otimes \mathrm{L}^{2}
(\mathbb{R})$:
\begin{equation} \label{eq:2}
d\ket{\Psi_{t}} = \left[ - \frac{i}{\hbar}\, H_{t}\, dt  +
\sqrt{\lambda} \left(q - \braket{q}_{t}\right) dW_{t} -
\frac{\lambda}{2} \left(q - \braket{q}_{t}\right)^{2} dt \right]
\ket{\Psi_{t}},
\end{equation}
which is precisely of the form~\eqref{eq:gen}, where~$H_{t}$ is the
(time dependent) standard Hamiltonian operator of the composite
system, and $q$~is the position operator associated to the centre of
mass of the pointer.\footnote{\label{fn:3}Thus, strictly speaking,
we should write~$I_{{\mathcal S}} \otimes q$ for the position
operator for the pointer, where~$I_{{\mathcal S}}$ is the identity
operator in~$\mathcal{H}_{{\mathcal S}}$.  We avoid such a way of
writing, when no confusion arises.} In the following we will use
capital letters ($\Psi, \Phi, \ldots$) to denote a state vector for
the composite ${\mathcal S} + {\mathcal A}$ system, and lower case
letters ($\psi, \phi, \ldots$) to denote a state vector referring to
the pointer alone.

As discussed in~\cite{ab1} we take for the constant $\lambda$
appearing in~\eqref{eq:2}
\begin{equation}
\label{eq:3} \lambda \simeq \frac{m}{m_{0}}\, \lambda_{0},
\end{equation}
with $m_{0} \simeq 1.7 \times 10^{-27}$ Kg being a reference mass of
order of a nucleon mass and $\lambda_{0} \simeq 10^{-2}$ m$^{-2}$
sec$^{-1}$.  For definiteness, let us consider a pointer of mass $m
= 1$ g (i.e., a pointer made of an Avogadro number of nucleons), and
let us define, for later convenience, the quantities
\begin{equation} \label{eq:4}
\omega := 2 \sqrt{\frac{\hbar \lambda}{m}} \; \simeq \; 5.0 \times
10^{-5}\,\makebox{sec$^{-1}$} \qquad\text{and}\qquad \sigma_{q} :=
\sqrt{\frac{\hbar}{m\omega}} \; \simeq \; 4.6 \times
10^{-14}\,\makebox{m}.
\end{equation}
% so that we have $\lambda = \omega / (4 \sigma_{q}^{2})$.

We take the Hamiltonian~$H_{t}$ to be of the form $H_{t} =
H_{{\mathcal S}} + H_{{\mathcal A}} + H_{\mathrm{INT}}$.  The first
term is the quantum Hamiltonian for the microscopic system: we
assume that the time scale of the free evolution of the microscopic
system is much larger than the characteristic time scale of the
experiment (``instantaneous measurement'' assumption); accordingly
we take~$H_{{\mathcal S}}$ to be the null operator. The second term
is the quantum Hamiltonian of the pointer, which we take equal to
that of a non-relativistic free quantum particle of mass~$m$:
$H_{{\mathcal A}} = p^{2} / (2m)$, where~$p$ is the momentum
operator.

Finally, we assume the interaction term~$H_{\mathrm{INT}}$ between
the two systems to be of the von~Neumann type, and devised in such a
way to measure the spin operator $S_{z}$:
\begin{equation} \label{eq:5}
H_{\mathrm{INT}}(t) = \kappa\, \Delta^{T}_{t}\, S_{z} \otimes p,
\end{equation}
where~$\kappa$ is a coupling constant and $\Delta^{T} \colon t
\mapsto \Delta^{T}_{t}$ is a $T$-normalized,\footnote{\label{fn:4}By
a $T$-normalized function, we just mean
\begin{equation*}
\int_{-\infty}^{+\infty} \Delta^{T}_{t}\, dt = \int_{t_{0}}^{t_{0} +
 T} \Delta^{T}_{t}\, dt = T.
\end{equation*}
Note that $\Delta^{T}_{t}$ depends also on the initial time $t_{0}$;
we will avoid to indicate it explicitly, when non confusion arises.}
non negative, real valued, function of time, identically equal to
zero outside a given interval of the form~$(t_{0}, t_{0} + T)$,
i.e., outside the time interval of length~$T$, say $T = 1$ sec,
during which the experiment takes place; we choose the time origin
in such a way that the experiment begins at $t_{0} = 0$ sec. As it
is well known, $H_{\mathrm{INT}}$ generates the following type of
evolution, depending on the initial state of the micro-system
${\mathcal S}$:
\begin{equation} \label{eq:6}
\left[ c_{+} \ket{+} + c_{-} \ket{-} \right] \otimes \ket{\phi_{0}}
\; \mapsto \;  c_{+} \ket{+} \otimes \ket{\phi_{+}}  + c_{-} \ket{-}
\otimes \ket{\phi_{-}},
\end{equation}
where $\ket{\phi_{\pm}}$ are final pointer states spatially
translated with respect to the initial state~$\ket{\phi_{0}}$ by the
quantity $\pm(\hbar/2)\, \kappa\, T$.

The strength of the coupling constant~$\kappa$ has to be chosen in
such a way that the distance $\hbar\, \kappa\, T$ between the
initial state $\ket{\phi_{0}}$ of the pointer and any of the two
final states $\ket{\phi_{\pm}}$ is macroscopic; for definiteness,
let us choose $\hbar\,\kappa = 1$ cm sec$^{-1}$, so that
$\hbar\,\kappa\,T = 1$ cm.

\subsection{The initial state}
\label{sec:initial-state} We take the initial states of the
microscopic system~${\mathcal S}$ and of the macroscopic
apparatus~${\mathcal A}$ to be completely uncorrelated, as it is
customary and appropriate for the description of a measurement
process. Accordingly, we assume the initial state of the total
system ${\mathcal S} + {\mathcal A}$ to be
\begin{equation} \label{eq:7}
\left[ c_{+} \ket{+} + c_{-} \ket{-} \right] \otimes \ket{\phi_{0}},
\end{equation}
where~$\ket{\phi_{0}}$ describes the ``ready'' state of the
macroscopic apparatus~${\mathcal A}$.

Regarding the initial state~$\ket{\phi^{{\mathcal A}}_{0}}$ of the
pointer, some considerations have to be done.  In~\cite{ab1} it has
been shown that, according to equation~\eqref{eq:2}, the wave
function for the centre of mass of an isolated quantum system
reaches asymptotically (and very rapidly, for a macro-object) a
Gaussian state of the form
\begin{equation} \label{eq:8}
\phi^{\mathrm{G}}_{t} (x) = \sqrt[4]{\frac{1}{2 \pi
\sigma_{q}^{2}}}\; \exp \left[ - \frac{1 - i}{4 \sigma_{q}^2} (x -
\bar{x}_{t})^2 + i\, \bar{k}_{t}\, x \right],
\end{equation}
(modulo a time-dependent global phase factor) with $\sigma_{q}$
defined as in Eq.~\eqref{eq:4}. For later reference, let us observe
that the dispersion of the Gaussian function of Eq.~\eqref{eq:8} in
momentum space is
\begin{equation} \label{eq:9}
\sigma_{p} = \frac{\hbar}{\sqrt{2}\, \sigma_{q}} \; \simeq \; 1.6
\times 10^{-21}\, \makebox{kg m sec$^{-1}$},
\end{equation}
quite close to the minimum allowed by Heisenberg's uncertainty
relation, and that the centres of~$\phi^{\mathrm{G}}_{t}$ in
position and momentum space are given by $\bar{x}_{t}$ and~$\hbar
\bar{k}_{t}$ respectively.

In our measurement model, we assume that the pointer is isolated for
the time prior to the experiment; during this time, as shown in the
past literature, its wave function converges rapidly towards a state
close to~\eqref{eq:8}, which we therefore assume to be the initial
state of the pointer. To summarize, we take as the initial state of
the composite system ${\mathcal S} + {\mathcal A}$ the ket
\begin{equation} \label{eq:10}
\ket{\Psi_{0}} = \left[ c_{+} \ket{+} + c_{-} \ket{-} \right]
\otimes \ket{\mathrm{G}, 0},
\end{equation}
where~$\braket{x | \mathrm{G}, 0}$ is of the form~\eqref{eq:8}. We
choose the natural reference frame where the pointer is initially at
rest, so that $\bar{k}_{0} = 0$ m$^{-1}$, with the origin set up in
such a way that $\bar{x}_{0} = 0$ m.

\section{Measurement of an eigenstate}
\label{sec:eigenstate-measurement}
We begin our study of the model by looking for the solution of
equation~\eqref{eq:2} satisfying the initial condition
\begin{equation} \label{eq:11}
\ket{\Psi^{\pm}_{0}} = \ket{\pm} \otimes \Ket{\mathrm{G}, 0},
\end{equation}
where the symbol~$\pm$ means that the state~$\ket{\pm}$ is {\it
either}~$\ket{+}$ {\it or}~$\ket{-}$, i.e., an eigenstate of the
operator~$S_{z}$.  We will show that, in this special case, the
state of the microscopic system does not change in time, while the
pointer moves along the scale so to give the correct outcome of the
measurement.

\subsection{The linear equation}
\label{sec:linear-equation} Following the standard procedure
outlined e.g. in~\cite{ab1} we pass from the non-linear
equation~\eqref{eq:2} to the corresponding linear equation
\begin{equation} \label{eq:12}
d\ket{\Phi_{t}} = \left[ - \frac{i}{\hbar}\, H_{t}\, dt +
\sqrt{\lambda}\, q\,  d\xi_{t} - \frac{\lambda}{2}\, q^{2}\, dt
\right] \ket{\Phi_{t}},
\end{equation}
where $\xi_{t}$ is a standard Wiener process defined on the
probability space $(\Omega, \mathcal{F}, \mathbb{Q})$, the measure
$\mathbb{Q}$~being a new probability measure chosen in such a way
that the old probability measure~$\mathbb{P}$ is generated from
$\mathbb{Q}$ by the martingale $\| \Phi_{t} \|^2$.  The Wiener
process~$\xi_{t}$ of equation~\eqref{eq:12} is related to the Wiener
process~$W_{t}$ of equation~\eqref{eq:2} via Girsonov's
rule~\cite{ls}:
\begin{equation} \label{eq:22}
dW_{t} = d\xi_{t} - 2 \sqrt{\lambda} \braket{q}_{t} dt.
\end{equation}
It is easy to prove that a vector of the form
\begin{equation} \label{eq:13}
\ket{\Phi^{\pm}_{t}} = \ket{\pm} \otimes \Ket{\phi^{\pm}_{t}}
\end{equation}
solves equation~\eqref{eq:12}, for the initial
condition~\eqref{eq:11}, if the wave function $\phi^{\pm}_{t} (x) :=
\braket{x | \phi^{\pm}_{t}}$ solves the following linear SDE, which
involves the apparatus degrees of freedom alone:
\begin{equation} \label{eq:14}
d\phi^{\pm}_{t}(x) \; = \; \left[ \left(
\frac{i\hbar}{2m}\,\frac{d^{2}}{dx^{2}} \mp \frac{\hbar\,
\kappa}{2}\, \Delta^{T}_{t}\, \frac{d}{dx} \right)\, dt \right.
\left. \; + \; \sqrt{\lambda}\, x \, d\xi_{t} \; - \;
\frac{\lambda}{2}\, x^2 \, dt \right] \phi^{\pm}_{t}(x ).
\end{equation}

\subsection{The solution and its properties}
\label{sec:solution-properties} The solution of
equation~\eqref{eq:14} for the given initial condition is the
Gaussian wave function
\begin{equation} \label{eq:15}
\phi^{\pm}_{t} (x) = \exp \left[ -  \alpha_{t} (x -
\bar{x}^{\pm}_{t})^{2} + i\, \bar{k}^{\pm}_{t}\, x  +
\gamma^{\pm}_{t} + i\, \theta^{\pm}_{t} \right],
\end{equation}
whose parameters $\alpha_{t} \in \mathbb{C}$, and
$\bar{x}^{\pm}_{t}, \bar{k}^{\pm}_{t}, \gamma^{\pm}_{t},
\theta^{\pm}_{t} \in \mathbb{R}$ (of obvious meaning) satisfy the
following system of SDE~\cite{ab1}:
\begin{eqnarray}  \label{eq:16}
d\alpha_{t} & = & \left( \lambda - \frac{2i\hbar}{m}\, \alpha_{t}^2
\right)\, dt \\
\label{eq:17} d\bar{x}^{\pm}_{t} & = & \left( \frac{\hbar}{m}\,
\bar{k}^{\pm}_{t} \pm \frac{\hbar}{2}\, \kappa \, \Delta^{T}_{t}
\right) dt \; + \; \frac{\sqrt{\lambda}}{2 \alpha^{\makebox{\tiny
R}}_{t}} \left\{ d\xi_{t} - 2\sqrt{\lambda}\, \bar{x}^{\pm}_{t}\, dt
\right\}\\
\label{eq:18} d\bar{k}^{\pm}_{t} & = & - \sqrt{\lambda}\,
\frac{\alpha^{\makebox{\tiny I}}_{t}}{\alpha^{\makebox{\tiny
R}}_{t}} \left\{ d\xi_{t} - 2\sqrt{\lambda}\, \bar{x}^{\pm}_{t}\, dt
\right\}\\
\label{eq:19} d\gamma^{\pm}_{t} & = & \left( \lambda
(\bar{x}^{\pm}_{t})^{2} + \frac{\hbar}{m}\, \alpha^{\makebox{\tiny
I}}_{t} + \frac{\lambda}{4 \alpha^{\makebox{\tiny R}}_{t}}\right)dt
\; + \;  \sqrt{\lambda}\, \bar{x}^{\pm}_{t} \left\{ d\xi_{t} -
2\sqrt{\lambda}\, \bar{x}^{\pm}_{t}\, dt
\right\}\\
\label{eq:20} d\theta^{\pm}_{t} & = & \left( - \frac{\hbar}{2m}\,
(\bar{k}^{\pm}_{t})^{2} - \frac{\hbar}{m}\, \alpha^{\makebox{\tiny
R}}_{t} + \frac{\lambda \alpha^{\makebox{\tiny I}}_{t}}{4
(\alpha^{\makebox{\tiny R}}_{t})^2} \mp \frac{\hbar \kappa}{2}\,
\Delta_{T}(t)\, \bar{k}^{\pm}_{t} \right)dt +  \sqrt{\lambda}
\frac{\alpha^{\makebox{\tiny I}}_{t}}{\alpha^{\makebox{\tiny
R}}_{t}}\,\bar{x}^{\pm}_{t} \left\{ d\xi_{t} - 2\sqrt{\lambda}\,
\bar{x}^{\pm}_{t}\, dt \right\}, \;\;\;\;\;
\end{eqnarray}
where we have denoted by~$z^{\makebox{\tiny R}}$ ($z^{\makebox{\tiny
I}}$) the real (imaginary) part of the complex number~$z$.

Eqs.~\eqref{eq:19} and \eqref{eq:20} are of no particular interest
in this simple situation, because they just describe the time
evolution of the irrelevant norm and global phase of the Gaussian
solution.  Eq.~\eqref{eq:16} is independent of Eqs.
\eqref{eq:17}--\eqref{eq:20}; it is deterministic, and easily solved
by separation of variables:
\begin{equation} \label{eq:21}
\alpha_{t} = \frac{1-i}{4 \sigma_{q}^{2}}\, \tanh\left(\frac{\omega
t}{1 - i} + c_{0}\right),
\end{equation}
where $c_{0}$ sets the initial condition.  Eq.~\eqref{eq:21}
determines the time evolution of the spread in position and momentum
of the Gaussian wave function. In our case, given the initial
condition~$\alpha_{0} = (1-i)/4 \sigma_{q}^{2}$, it follows that
$\alpha_{t} \equiv (1-i)/4 \sigma_{q}^{2}$ for all times (i.e., we
can set $\alpha^{\makebox{\tiny R}}_{t} \equiv 1/4 \sigma_{q}^{2}$
and $\alpha^{\makebox{\tiny I}}_{t} \equiv -1 /4 \sigma_{q}^{2}$ in
equations~\eqref{eq:17}--\eqref{eq:20}): as expected, the position
and momentum spreads of the wave function of the pointer do not
change in time, and remain identically equal to, respectively,
$\sigma_{q}$ and $\sigma_{p}$.

For what concerns equations~\eqref{eq:17} and \eqref{eq:18}, which
do not depend on equations~\eqref{eq:19} and \eqref{eq:20}, their
solution describes the time evolution of the mean value in position
and momentum of the Gaussian wave function.  We have to characterize
the stochastic properties of the solution with respect to the
physical probability measure~$\mathbb{P}$, i.e. we need to go back
to the original noise $W_{t}$ via Girsanov's rule~\eqref{eq:22},
which in this case is very easy, because for a wave function
like~\eqref{eq:15} we simply have $\braket{q}_{t} =
\bar{x}^{\pm}_{t}$ for all times, so that all we have to do is to
write~$dW_{t}$ in place of the curly braces~$\{\dots\}$ in
equations~\eqref{eq:17} and \eqref{eq:18}:
\begin{eqnarray} \label{eq:a1}
d\bar{x}^{\pm}_{t} & = & \left( \frac{\hbar}{m}\,\bar{k}^{\pm}_{t}
\pm \frac{\hbar\,\kappa}{2}\, \Delta^{T}_{t} \right)\, dt
+ \sigma_{q}\, \sqrt{\omega}\, dW_{t}\\
\label{eq:a2} d\bar{k}^{\pm}_{t} & = &
\frac{\sigma_{p}}{\sqrt{2}\hbar} \sqrt{\omega}\, dW_{t},
\end{eqnarray}
where we have also taken into account that $\alpha_{t} = (1-i)/4
\sigma_{q}^{2}$ for any $t \geq 0$. Let us call
$\ket{\Psi^{\pm}_{t}}$ the normalized physical solutions:
$\ket{\Psi^{\pm}_{t}} \equiv \ket{\Phi^{\pm}_{t}}/\|
\ket{\Phi^{\pm}_{t}} \|$, with $\ket{\Phi^{\pm}_{t}}$ given by
Eq.~\eqref{eq:13}; taking also into account that $\bar{x}_{0} = 0$
m, and $\bar{k}_{0} = 0$ m$^{-1}$, we find the following results.
\begin{enumerate}
\item\label{item:7} According to Eq.~\eqref{eq:a2}, the average value of the
peak of the Gaussian wave function in momentum space,
$\braket{p}^{\pm}_{t} := \langle \Psi^{\pm}_{t}| p | \Psi^{\pm}_{t}
\rangle \equiv \hbar \bar{k}^{\pm}_{t}$, does not evolve in time:
\begin{equation} \label{eq:23}
\mathbb{E}_{\mathbb P} [ \braket{p}^{\pm}_{t} ] = \bar{p}^{\pm}_{0}
= 0\; \makebox{Kg m sec$^{-1}$}.
\end{equation}
\item\label{item:8} By Eqs.~\eqref{eq:a1} and \eqref{eq:a2}, the average
value of the peak in position space $\braket{q}^{\pm}_{t} := \langle
\Psi^{\pm}_{t}| q | \Psi^{\pm}_{t} \rangle \equiv \bar{x}^{\pm}_{t}$
of the Gaussian wave function evolves in time according to
\begin{equation} \label{eq:24}
\mathbb{E}_{\mathbb P} [ \braket{q}^{\pm}_{t} ] \; = \;  \pm
\frac{\hbar\, \kappa}{2} \int_{0}^{t} \Delta^{T}_{t'}\, dt'.
\end{equation}
Equation~\eqref{eq:24} shows that, when the measurement begins,
$\mathbb{E}_{\mathbb P} [ \braket{q^{\pm}_{t}} ]$ moves towards the
right or the left according to the initial state of the microscopic
system~${\mathcal S}$, reaching the final value (at the end of the
measurement)
\begin{equation} \label{eq:25}
\mathbb{E}_{\mathbb P} [ \braket{q}^{\pm}_{t} ] \; = \; \pm
\frac{\hbar\, \kappa\, T}{2} \; = \; \pm 0.5\, \makebox{cm}
\qquad\text{for~$t \geq T$}.
\end{equation}
\item\label{item:9} The variance $\mathbb{V}_{\mathbb P} [
\braket{q}^{\pm}_{t} ] \equiv \mathbb{E}_{\mathbb P} [
\braket{q}^{\pm}_{t} - \mathbb{E}_{\mathbb P} [ \braket{q}^{\pm}_{t}
]]^2 $ associated to the motion of $\braket{q}^{\pm}_{t}$ is equal
to the variance computed in~\cite{ab1} (see Sec. VII B), which, for
$\alpha_{t} \equiv (1-i)/4 \sigma_{q}^{2}$, is given by\footnote{We
correct in this way a typo contained in Eq.~(93) of
ref.~\cite{ab1}.}:
\begin{equation} \label{eq:26}
\mathbb{V}_{\mathbb P} [ \braket{q}^{\pm}_{t} ] \; = \;
\sigma_{q}^{2} \left[ \omega t + \frac{(\omega t)^{2}}{2} +
\frac{(\omega t)^{3}}{12} \right];
\end{equation}
with our choices for the parameters, we have~$\mathbb{V}_{\mathbb P}
[ \braket{q}^{\pm}_{t} ] \leq \mathbb{V}_{\mathbb P} [
\braket{q}^{\pm}_{T} ] \simeq 1.1 \times 10^{-31}$ m$^{2}$, for any
$t \leq T$.
\end{enumerate}

From the above results we can derive the following important
conclusions:
\begin{itemize}
\item Due to the smallness of its variance,
the motion of the peak $\braket{q}^{\pm}_{t}$ of the
Gaussian wave function for the c.m. of the pointer is practically
{\it deterministic} and equivalent to the motion of
$\mathbb{E}_{\mathbb P} [ \braket{q}^{\pm}_{t} ]$, the fluctuations
around the mean being so tiny that they can be safely ignored. E.g.,
the probability for $\braket{q}^{\pm}_{t}$ to lie outside an
interval of width $\Delta$ centered in $\mathbb{E}_{\mathbb P} [
\braket{q}^{\pm}_{t} ]$ can be estimated by using \v{C}ebi\v{c}ev's
inequality; for $\Delta = 10^{-5}$ cm, we have
\begin{equation} \label{eq:27}
\mathbb{P}[ | \braket{q}^{\pm}_{t} -  \mathbb{E}_{\mathbb P} [
\braket{q}^{\pm}_{t} ]| \geq \Delta/2] \; \leq \; 4\,
\frac{\mathbb{V}_{\mathbb P} [ \braket{q^{\pm}_{T}} ]}{\Delta^{2}}
\; \simeq \; 4.2 \times 10^{-17}, \qquad \makebox{for any $t \leq
T$,}
\end{equation}
a vanishingly small probability.

\item As such, and because of Eq.~\eqref{eq:24},
the peak $\braket{q}^{\pm}_{t}$ evolves in time as follows:
\begin{equation} \label{eq:n1}
\braket{q}^{\pm}_{t} = \left\{
\begin{array}{lll}
\displaystyle \pm \frac{\hbar\, \kappa}{2} \int_{0}^{t}
\Delta^{T}_{t'}\, dt' &
t \leq T & \qquad \makebox{(+ negligible fluctuations)}\\
& & \\
\displaystyle \pm \frac{\hbar\, \kappa\, T}{2} \; = \; \pm 0.5\,
\makebox{cm} \quad & t \geq T & \qquad \makebox{(+ negligible
fluctuations)}
\end{array}
\right.
\end{equation}
This means that, according to the initial state of the micro-system,
the pointer moves in a practically deterministic way either towards
the left or towards the right, with respect to the initial
ready-state, displaying in this way the outcome of the measurement.

\item During the measurement, the state of the micro-system does
not change.
\end{itemize}
This is precisely the expected behavior both for the microscopic
system as well as for the macroscopic pointer, when the initial
state is given by~\eqref{eq:13}, for an ideal measurement scheme as
the one of von Neumann here analyzed.

\section{Measurement of a superposition}
\label{sec:superposition-measurement}

Let us now consider the general case where the initial
state~$\ket{s_{i}}$ of the microscopic system~$\mathcal{S}$ is not
an eigenstate of~$S_{z}$, but a superposition of eigenstates of the
form
\begin{equation}
  \label{eq:28}
  \ket{s_{i}} = c_{+} \ket{+} + c_{-} \ket{-}
  \qquad \text{($|c_{+}|^{2} + |c_{-}|^{2} = 1$)};
\end{equation}
the global initial condition for the micro-system and the apparatus
then is
\begin{equation}
  \label{eq:29}
  \ket{\Psi_{0}} \quad = \quad \left[
    c_{+} \ket{+} + c_{-} \ket{-}
  \right] \otimes \ket{G, 0}.
\end{equation}
As in the preceding subsection, we first solve the linear
equation, and next move to the non-linear one.  Due to the
linearity of equation~\eqref{eq:12}, its solution, with the given
initial condition~\eqref{eq:29}, is
\begin{equation}
  \label{eq:30}
  \ket{\Phi_{t}} = \ket{+} \otimes \ket{\phi^{+}_{t}}
  + \ket{-} \otimes \ket{\phi^{-}_{t}},
\end{equation}
where the wave functions~$\ket{\phi^{+}_{t}}$
and~$\ket{\phi^{-}_{t}}$, in the position representation, are of the
form~\eqref{eq:15} and the parameters $\alpha_{t},
\bar{x}^{\pm}_{t}, \bar{k}^{\pm}_{t}, \gamma^{\pm}_{t},
\theta^{\pm}_{t}$ solve Eqs.~\eqref{eq:16} to~\eqref{eq:20}, with
the obvious choice of sign and with initial conditions:
\begin{equation}
\alpha_{0} = \frac{1-i}{4\sigma_{q}^2}, \qquad \bar{x}^{\pm}_{0} =
0\; \makebox{m}, \qquad \bar{k}^{\pm}_{0} = 0\; \makebox{m$^{-1}$},
\qquad \gamma^{\pm}_{0} = \ln | c_{\pm} |, \qquad \theta^{\pm}_{0} =
\makebox{Arg}[c_{\pm}]
\end{equation}
(of course we now assume that $c_{\pm} \neq 0$).

Since the time evolution of the parameters~$\alpha_{t}$ is governed
by equation~\eqref{eq:16} which, as we have already remarked, is
deterministic and decoupled from the evolution
equations~\eqref{eq:17}--\eqref{eq:20} for the remaining parameters,
we observe first of all that the evolution of the spreads in
position and momentum of the two Gaussian
functions~$\ket{\phi^{+}_{t}}$ and~$\ket{\phi^{-}_{t}}$ does not
change with respect to the case analyzed in the previous section.
Accordingly, we have $\alpha_{t} \equiv (1-i)/4 \sigma_{q}^{2}$ for
all times, so that the spreads of the two wave functions do not
evolve and remain identically equal to the asymptotic
values~$\sigma_{q}$ and~$\sigma_{p}$.

\subsection{The deterministic evolution of the distances in position and
  momentum between the two Gaussian components}
\label{sec:gaussians-distances}

Contrary to the preceding case, moving from the solution of the
linear equation to the solution of the non-linear one is not
immediate, since Girsanov's rule~\eqref{eq:22} involves the quantum
average~$\braket{q}_{t}$, which in this case turns out not to be a
trivial function of the parameters controlling the two Gaussian
components; namely, one finds that\footnote{In Eq.~\eqref{eq:vmdg}
no contribution comes from the overlapping between the two Gaussian
components, since each component is coupled to one of the two
orthogonal spin state $|\pm\rangle$, which make the ``off-diagonal''
terms of the scalar product vanish.}:
\begin{equation} \label{eq:vmdg}
\braket{q}_{t} \; = \; \frac{\bar{x}^{+}_{t} e^{2 \gamma^{+}_{t}} +
\bar{x}^{-}_{t} e^{2 \gamma^{-}_{t}}}{e^{2 \gamma^{+}_{t}} + e^{2
\gamma^{-}_{t}}}.
\end{equation}
Of course, this is an entirely expected difficulty, due to the
essential non-linearity inherent to collapse models; to proceed in
the analysis of the problem, it is convenient first of all to
analyze the evolution of the distance between the maxima of the two
Gaussian functions~$\ket{\phi^{+}_{t}}$ and~$\ket{\phi^{-}_{t}}$,
both in position as well as in momentum space, and subsequently of
their relative weights.

Following the path outlined in~\cite{ab1}, let us consider the
differences $X_{t} := \bar{x}^{+}_{t} - \bar{x}^{-}_{t}$ and $K_{t}
:= \bar{k}^{+}_{t} - \bar{k}^{-}_{t}$, which express at each instant
the distance in position and (modulus $\hbar$) momentum space
between the centres of the two Gaussian
functions~$\ket{\phi^{+}_{t}}$ and~$\ket{\phi^{-}_{t}}$.  From
equations~\eqref{eq:17} and \eqref{eq:18}, keeping in mind that in
our case we have $\alpha^{\makebox{\tiny R}}_{t} \equiv 1/4
\sigma_{q}^{2}$ and $\alpha^{\makebox{\tiny I}}_{t} \equiv -1/4
\sigma_{q}^{2}$ for all times, we get the following
\emph{deterministic} system for~$X_{t}$ and~$K_{t}$:
\begin{equation}
  \label{eq:31}
  \frac{d}{dt}
  \begin{bmatrix}
    X_{t} \\ K_{t}
  \end{bmatrix} =
  \begin{bmatrix}
    - \omega                  & \hbar/m\\
    - 2 \lambda & 0
  \end{bmatrix}
  \begin{bmatrix}
    X_{t} \\ K_{t}
  \end{bmatrix} +
  \begin{bmatrix}
    \hbar\, \kappa\, \Delta^{T}_{t}\\ 0
  \end{bmatrix};
\end{equation}
since it does not depend on the noise, it is insensitive to the
change of measure and holds true also for the non-linear
Eq.~\eqref{eq:2}.

The solution of the above system depends of course on the specific
choice for the function $\Delta^{T}_{t}$; a simple reasonable choice
is the following:
\begin{equation}
\Delta^{T}_{t} \quad = \quad \left\{
\begin{array}{ll}
1 \quad & t \, \in \, [0,T] \\
0 & \makebox{else},
\end{array}
\right.
\end{equation}
which, according to Eq.~\eqref{eq:n1}, means that, with restriction
to the situation analyzed in the previous section, during the
measurement the pointer moves at a constant speed either towards the
left or towards the right, depending on the initial state of the
micro-system. According to this choice, $X_{t}$, given the initial
condition $X_{0} = 0$ m, evolves in time as follows:
\begin{equation}
  \label{eq:n2}
  X_{t} \; = \;
  \left\{
  \begin{array}{ll}
  \displaystyle
    \frac{2\hbar\kappa}{\omega}\,
    e^{-\omega t/2}
    \sin\frac{\omega}{2}\, t \quad &
    \text{for $0\leq t\leq T$},  \\
  & \\
  \displaystyle
  \frac{2\hbar\kappa}{\omega}\,
    e^{-\omega t/2}\left[
    \sin\frac{\omega}{2}\, t
    \, - \, e^{\omega T/2}
    \sin\frac{\omega}{2}\, (t-T)
    \right] \quad &
    \text{for $t \geq T$}.
 \end{array}
 \right.
\end{equation}
Since $\omega^{-1} \simeq 2.0 \times 10^{4}$ sec is a very long time
compared to the measurement-time, we can meaningfully expand
Eq.~\eqref{eq:n2} to first order in $\omega t$:
\begin{equation}
  \label{eq:n2bis}
  X_{t} \; \simeq \;
  \left\{
  \begin{array}{ll}
  \hbar \kappa\, t \quad\qquad &
    \text{for $0\leq t\leq T = (\hbar\kappa)^{-1}$} \; = \; 1\, \text{sec},  \\
  1 \; \makebox{cm} \quad &
    \text{for $T \leq t \ll \omega^{-1} \simeq 2.0 \times 10^{4}$ sec}
 \end{array}
 \right.
\end{equation}
As we see, the distance between the two peaks increases almost
linearly in time, reaching its maximum (1 cm) at the end of the
measurement process, as predicted by the standard Schr\"odinger
equation; after this time, their distance remains practically
unaltered for extremely long times, and only for $t \simeq 2.0
\times 10^{4}$ sec it starts slowly decreasing, eventually going to
0. Note that such a behavior, being determined by $\omega$, does
{\it not} depend on the mass of the pointer, thus a larger pointer
will not change the situation. The moral is that $X_{t}$ behaves as
if the reduction mechanism were not present (as if $\lambda_{0} =
0$) so we have to look for the collapse somewhere else.

As we shall discuss in the next subsection, the collapse occurs
because, in a very short time, the weight of one of the two Gaussian
wave functions ($\ket{\phi^{+}_{t}}$ or $\ket{\phi^{-}_{t}}$)
becomes much smaller than the weight of the other component; this
implies that, when the normalization of the whole state is taken
into account, one of the two components practically disappears, and
only the other one survives, the one which sets the outcome of the
experiment. Of course, this process is random and, as we shall
prove, it occurs with a probability almost equivalent to the Born
probability rule.

\subsection{The evolution equation governing the relative weight of the
  two Gaussian components}
\label{sec:gaussians-weights}

The relative damping between the two Gaussian components of
Eq.~\eqref{eq:30} is measured by the stochastic process $\Gamma_{t}
= \gamma^{+}_{t} - \gamma^{-}_{t}$: if, at a certain time $t$, it
occurs that $\Gamma_{t} \gg 1$, it means that at the end of the
experiment~$\ket{\phi^{-}_{t}}$ is suppressed with respect
to~$\ket{\phi^{+}_{t}}$, so that the initial state~\eqref{eq:29}
practically evolves to~$\ket{+} \otimes \ket{\psi^{+}_{t}}$
(remember that $\ket{\psi^{\pm}_{t}} = \ket{\phi^{\pm}_{t}}/\|
\ket{\phi^{\pm}_{t}}\|$ are the normalized states); the opposite
happens if $\Gamma_{t} \ll -1$. To be quantitative, let us introduce
a conveniently large collapse parameter, say 35, and the following
definition\footnote{The choice made here for the collapse parameter
is different from the one made in~\cite{ab1}. We find this new
choice, which at any rate is arbitrary, more convenient for the
problem under study.}:
\begin{quote}
{\bf Definition.} The superposition~\eqref{eq:30} is {\it
suppressed} when $|\Gamma_{t}| \geq 35$, i.e., when either
$\|\ket{\phi^{+}_{t}}\| / \|\ket{\phi^{-}_{t}}\|$ or its reciprocal
is greater than $e^{35} \simeq 1.6 \times 10^{15}$.
\end{quote}
Using equation~\eqref{eq:19} and the Girsonov's
transformation~\eqref{eq:22}, we can write the following SDE
for~$\Gamma_{t}$ in terms of the noise $W_{t}$ associated to the
non-linear Eq.~\eqref{eq:2}:
\begin{equation}
  \label{eq:37}
  d \Gamma_{t} \; = \; \lambda X_{t} \left(
    2 \braket{q}_{t} - \bar{x}^{+}_{t} - \bar{x}^{-}_{t}
  \right) dt
  + \sqrt{\lambda} X_{t}\, dW_{t},
\end{equation}
with initial condition $\Gamma_{0} = \ln |c_{+}/c_{-}|$. By using
the expression~\eqref{eq:vmdg} for $\braket{q}_{t}$, we can re-write
the above equation as follows:
\begin{equation}
  \label{eq:37b}
  d \Gamma_{t} \; = \; \lambda X_{t}^{2} \tanh \Gamma_{t}\, dt
  \, + \, \sqrt{\lambda} X_{t}\, dW_{t}.
\end{equation}
This is the result we wanted to arrive at, and we will devote the
rest of the section at analyzing its physical content. To proceed
further with the analysis, it is convenient to perform the following
time change~\cite{gs},
\begin{equation}
  \label{eq:45}
  t \quad \longrightarrow \quad s_{t} \; := \; \lambda \int_{0}^{t}
  X_{t}^{2}\, dt',
\end{equation}
which allows us to describe the collapse process in terms of the
dimensionless quantity $s$ that measures its effectiveness. Using
Eq.~\eqref{eq:n2}, one can solve exactly the above integral and
compute $s$ as a function of $t$. Such a function however cannot be
inverted in order to get $t$ from $s$. To this end, we use the
simplified expression~\eqref{eq:n2bis} in place of the the exact
formula Eq.~\eqref{eq:n2} to compute the integral, an expression
which, as we have seen, represents a very good approximation to the
time evolution of $X_{t}$ throughout the whole time during which the
experiment takes place (alternatively, we may initially choose
$\Delta^{T}_{t}$ in such a way that $X_{t}$ evolves exactly like
in~\eqref{eq:n2bis}, at least from $t = 0$ to $t = T$). Accordingly,
we have:
\begin{eqnarray} \label{eq:n45}
s \; \equiv \; s_{t} & \simeq & \frac{\lambda \hbar^2 \kappa^2}{3}\,
t^3 \; \simeq \; 2.0 \times 10^{17}\, (t/\text{sec})^3
\;\;\quad\qquad 0
\, \leq t \, \leq T = 1 \; \text{sec}, \\
t \; \equiv \; t_{s} & \simeq & \sqrt[3]{\frac{3}{\lambda \hbar^2
\kappa^2}\,s} \; \simeq \; (1.7 \times 10^{-6}\, \sqrt[3]{s})\;
\text{sec} \qquad\; 0 \, \leq s \, \leq \lambda \hbar^2 \kappa^2/3 =
2.0 \times 10^{17}.
% \label{eq:n5}
\end{eqnarray}
Note that, according to the above equations, the physical time $t$
depends on $s$ through the inverse cubic-root of $\lambda$, i.e. on
the inverse cubic-root of the mass of the pointer; this time
dependence of $t$ on $\lambda$ is important since, as we shall see,
it will affect the collapse time. We do not study the functional
dependence between $s$ and $t$ for $t \geq T$ since, as we shall
soon see and as we expect it to be, the collapse occurs at times
much smaller than $T$.

Written in terms of the new variable $s$, Eq.~\eqref{eq:37b} reduces
to:
\begin{equation} \label{eq:n14} d \Gamma_{s} \; = \; \tanh
\Gamma_{s} \, ds \, + \, dW_{s};
\end{equation}
this equation has been already analyzed in~\cite{ab1}, using the
theorems of~\cite{gs}; here we report the main properties.

\subsubsection{The Collapse Time}

According to the definition given before, a collapse occurs when
$|\Gamma_{t}| \geq 35$; it would seem then natural to define the
collapse time as the time when $|\Gamma_{t}|$ first reaches the
value 35. However, one has to face the event that $|\Gamma_{t}|$,
after having reached such a value, immediately starts decreasing in
a significant way, jeopardizing in this way the effect of the
collapse. To avoid such a possibility, we proceed as follows: we
will compute the time it takes for $|\Gamma_{t}|$ to reacher a value
larger than 35, let us say 50, and subsequently we will show that,
after having reached such a value, the probability that it gets back
to a value below 35 is negligible. In this way we can be (almost)
sure\footnote{Here, as well as in the rest of the paper, we use
``almost sure'' in the physical sense of ``with very high
probability'', not in the mathematical sense of ``with the possible
exception of a subset of measure 0''.} that, once the collapse has
occurred, the superposition never re-appears.

Let us consider the time $\bar{S} = \bar{S}(\omega)$ when
$|\Gamma_{s}|$ first reaches the value 50:
\begin{equation}
\bar{S} \; \equiv \; \inf \{ s: \;\; |\Gamma_{s}| \geq 50 \};
\end{equation}
of course we assume that the initial state~\eqref{eq:28} is such
that $|\Gamma_{0}| \leq 35$, otherwise according to our definition
(as well as for all practical purposes) it would already be a
reduced state, not a physically meaningful superposition. It can be
proven~\cite{ab1} that $\bar{S}$ is {\it finite} with probability 1,
and that its average value and variance are given by the following
expressions:
\begin{eqnarray}
{\mathbb E}_{\mathbb P}[\bar{S}] & = & 50 \tanh 50 \, - \,
\Gamma_{0}
\tanh \Gamma_{0}, \\
{\mathbb V}_{\mathbb P}[\bar{S}] & = & F(50) - F(\Gamma_{0}), \quad
F(x) = x^2 \tanh^2 x + x \tanh x - x^2.
\end{eqnarray}
Now, $\tanh 50 \simeq 1 - 7.4 \times 10^{-44}$ which is practically
1; let us also consider e.g. the worst case, as far as the collapse
mechanism is concerned, i.e. the case in which $\Gamma_{0} = 0$,
which means that the micro-state is initially in a equal weighted
superposition of the two eigenstates. We then have that ${\mathbb
E}_{\mathbb P}[\bar{S}] \simeq 50$ and ${\mathbb V}_{\mathbb
P}[\bar{S}] \simeq 50$.

$\bar{S}(\omega)$ is a random variable, so we can not tell exactly
when the collapse occurs; since however we want to be quite safe
that it actually occurs, let us compute the probability that
$\bar{S}$ happens to be much greater than, e.g., $10^5$ times its
standard deviation. By a trivial application of \v{C}ebi\v{c}ev's
inequality we have:
\begin{equation}
{\mathbb P}\left[ |\bar{S} - {\mathbb E}_{\mathbb P}[\bar{S}]| \geq
10^5 \sqrt{{\mathbb V}_{\mathbb P}[\bar{S}]} \; \simeq \; 7.1 \times
10^5 \right] \; \leq \; 10^{-10}.
\end{equation}

We can then conclude that, at time $s = {\mathbb E}_{\mathbb
P}[\bar{S}] + 10^5 \sqrt{{\mathbb V}_{\mathbb P}[\bar{S}]} \simeq
7.1 \times 10^5$, the collapse has almost certainly occurred (with
probability greater than $1 - 10^{-10}$) and that it is an
irreversible process (as we shall soon prove). Moving back from the
effective time $s$ to the physical times $t$ by using
Eq.~\eqref{eq:n45}, we then define the collapse time as follows:
\begin{equation} \label{eq:n13}
T_{C} \; \simeq \; \sqrt[3]{\frac{3({\mathbb E}_{\mathbb P}[\bar{S}]
+ 10^5 \sqrt{{\mathbb V}_{\mathbb P}[\bar{S}]}) }{\lambda \hbar^2
\kappa^2}} \; \simeq \; 1.5 \times 10^{-4} \, \makebox{sec}:
\end{equation}
the collapse occurs within a time interval smaller than the
perception time of a human observer. The above formula shows also
that, as expected, $T_{C}$ is proportional to the inverse cubic-root
of the mass of the pointer (since $\lambda$ is proportional to the
mass): the bigger the pointer, the shorter the collapse time. With
our choice for $\lambda_{0}$, even for a 1-g pointer the reduction
occurs practically instantaneously.

It is important to note that, at time $T_{C} \simeq 1.5 \times
10^{-4}$ sec, the distance between the two Gaussian components is
approximately $X_{T_{C}} \simeq 1.5 \times 10^{-4}$ cm: this means
that, with very high probability, the collapse occurs {\it before}
the two components have enough time to spread out in space to form a
macroscopic superposition. This means that, from the physical point
of view, there is {\it no} collapse of the wave function at all,
since it always remains perfectly localized in space at any stage of
the experiment. In any case, we will keep talking of collapse of the
wave function, meaning with it simply the event $|\Gamma_{t}| \geq
35$.

\subsubsection{The Collapse Probability}

Let us call $P_{+}$ the probability that $\Gamma_{s}$ hits the point
$+50$ before the point $-50$, i.e. the probability that
$|\phi^{+}_{s}\rangle$ survives during the collapse process so that
the outcome of the measurement is ``$+\hbar/2$''. Such a probability
turns out to be equal to~\cite{ab1}:
\begin{equation}
P_{+} \; = \; \frac{1}{2}\, \frac{\tanh 50 + \tanh \Gamma_{0}}{\tanh
50};
\end{equation}
while the probability $P_{-}$ that $\Gamma_{s}$ hits the point $-50$
before the point $+50$, i.e. that the outcome of the experiment is
``$-\hbar/2$'', is of course:
\begin{equation}
P_{-} \; = \; \frac{1}{2}\, \frac{\tanh 50 - \tanh \Gamma_{0}}{\tanh
50}.
\end{equation}
By taking into account that $\tanh 50 \simeq 1 - 7.4 \times 10^{-44}
\simeq 1$, we can write, with extremely good approximation:
\begin{eqnarray}
P_{+} & \simeq & \frac{1}{2}\, \left[ 1 + \tanh \Gamma_{0} \right]
\; = \; \frac{e^{\Gamma_{0}}}{e^{\Gamma_{0}} + e^{-\Gamma_{0}}} \; =
\; \frac{e^{2\gamma^{+}_{0}}}{e^{2\gamma^{+}_{0}} +
e^{2\gamma^{-}_{0}}} \; = \; |c_{+}|^2,
\\
P_{-} & \simeq & \frac{1}{2}\, \left[ 1 - \tanh \Gamma_{0} \right]
\; = \; \frac{e^{-\Gamma_{0}}}{e^{\Gamma_{0}} + e^{-\Gamma_{0}}} \;
= \; \frac{e^{2\gamma^{-}_{0}}}{e^{2\gamma^{+}_{0}} +
e^{2\gamma^{-}_{0}}} \; = \; |c_{-}|^2.
\end{eqnarray}
We see that the probability of getting one of the two possible
outcomes is practically {\it equivalent to the Born probability
rule!} On the one hand, this is an entirely expected results, since
collapse models have been designed precisely in order to solve the
measurement problem and in particular to reproduce quantum
probabilities; on the other hand, it is striking that a very general
equation like Eq.~\eqref{eq:2}, which is meant to describe both
quantum systems as well as macroscopic classical objects (i.e. all
physical situations, at the non relativistic level), when applied to
a measurement situation, provides not only a consistent description
of the measurement process, but also reproduces quantum
probabilities with such a good precision.

\subsubsection{Stability of the Collapse Process}

We have already anticipated that, since $\Gamma_{s}$ evolves
randomly, there is the chance that, after having reached e.g. the
value $+50$, i.e. after that the wave function collapsed to the
state $|\phi^{+}\rangle$, it becomes smaller than $50$ instead of
keeping increasing, eventually getting closer and closer to 0. When
such an event occurs, the superposition of the two Gaussian wave
functions, which was previously reduced, reappears again,
jeopardizing in this way the entire collapse process and
localization properties of the pointer. We now give an estimate of
the probability for such an event to occur.

Let us call $Q_{+}$ the probability that $\Gamma_{s}$, after having
reached the value $+50$ at time $\bar{S}$, does {\it not} go back to
a value smaller $35$:
\begin{equation}
Q_{+}\; := \; {\mathbb P}\left[ \inf_{s \geq \bar{S}} \Gamma_{s}
\geq 35 \right];
\end{equation}
such a probability turns out to be~\cite{ab1}:
\begin{equation}
Q_{+}\; \geq \; (1 + \tanh 50)\, \frac{\tanh 15}{1 + \tanh 15} \;
\simeq \; 1 - 9.3 \times 10^{-14},
\end{equation}
which is practically equal to 1: once a localization occurs, the
superposition can de facto never re-appear.

\subsection{State vector after the collapse}
\label{sec:sat-vec-aft-coll}

At time $t \geq T_{C}$ the normalized sate vector $|\Psi_{t}\rangle
\equiv |\Phi_{t}\rangle/\| |\Phi_{t}\rangle \|$, with
$|\Phi_{t}\rangle$ given in~\eqref{eq:30}, is:
\begin{equation}
|\Psi_{t}\rangle \quad = \quad \frac{|+\rangle \otimes |G +,
t\rangle \; + \; \epsilon_{t} |-\rangle \otimes |G -,
t\rangle}{\sqrt{1 + \epsilon_{t}^2}},
\end{equation}
where $\epsilon_{t} \; \equiv \; e^{-(\gamma^{+}_{t} -
\gamma^{-}_{t})} $ and the normalized Gaussian states $\langle x|G
\pm, t\rangle$ are defined as follows:
\begin{equation}
\langle x|G \pm, t\rangle \quad = \quad
\sqrt[4]{\frac{1}{2\pi\sigma_{q}^{2}}} \, \exp \left[ -
\frac{1-i}{4\sigma_{q}^{2}} (x - \bar{x}^{\pm}_{t})^{2} + i\,
\bar{k}^{\pm}_{t}\, x + i\, \theta^{\pm}_{t} \right].
\end{equation}
Let us assume that the collapse occurred in favor of the
``$+\hbar/2$'' eigenvalue, i.e. in such a way that $\Gamma_{t} \geq
35$ for $t \geq T_{C}$, with very high probability; it follows that:
\begin{equation}
\epsilon_{t} \; \leq \; e^{-35} \; \simeq \; 6.3 \times 10^{-16}
\qquad \forall \,\, t \geq T_{C},
\end{equation}
and we can write, with excellent accuracy:
\begin{equation}
|\Psi_{t}\rangle \quad \simeq \quad |+\rangle \otimes |G +,
t\rangle.
\end{equation}
We recover in this way the {\it postulate of wave packet reduction}
of standard quantum mechanics: at the end of the measurement
process, the state of the micro-system is reduced to the eigenstate
corresponding to the eigenvalue which has been obtained as the
outcome of the measurement, the outcome being defined by the
surviving Gaussian component ($|G +, t\rangle$ in this case). Note
the important fact that, according to our model, the collapse acts
directly only on the pointer of the measuring apparatus, not on the
micro-system; however, the combined effect of the collapse plus the
von Neumann type of interaction is that the microscopic
superposition of the spin states of the micro-system gets rapidly
reduced right after the measurement.

Note finally that, after the collapse, the states of the
micro-system and of the pointer are de facto factorized: as such,
after the measurement process one can, for all practical purposes,
disregard the pointer and focus only on the micro-system for future
experiments or interactions with other systems, as it is custom in
laboratories.

\subsection{The end of the experiment}
\label{sec:final-phase}

In this final subsection we study how, after the collapse, the
``winning'' component ($|G +, t\rangle$ or $|G -, t\rangle$) moves
in space, i.e. how their centers $\bar{x}^{+}_{t}$ or
$\bar{x}^{-}_{t}$ evolve in time, whether they move in such a way to
display the correct outcome of the measurement. To this purpose let
us define:
\begin{equation} \label{eq:cv}
\tilde{X}_{t} \; \equiv \; \bar{x}^{+}_{t} + \bar{x}^{-}_{t},
\qquad\quad \tilde{K}_{t} \; \equiv \; \bar{k}^{+}_{t} +
\bar{k}^{-}_{t},
\end{equation}
so that $\bar{x}^{+}_{t}$ and $\bar{x}^{-}_{t}$ as functions of
$X_{t}$ and $\tilde{X}_{t}$ are given by: $\bar{x}^{+}_{t} = (X_{t}
+ \tilde{X}_{t})/2$ and $\bar{x}^{-}_{t} = -(X_{t} -
\tilde{X}_{t})/2$. From Eqs.~\eqref{eq:17} and~\eqref{eq:18}, taking
also into account~\eqref{eq:vmdg}, one finds out that
$\tilde{X}_{t}$ and $\tilde{K}_{t}$ satisfy the following SDEs:
\begin{equation} \label{eq:nlinsys}
\begin{array}{lcll}
d \tilde{X}_{t} & = & \displaystyle \frac{\hbar}{m}\,\tilde{K}_{t}\,
dt \, + \, \omega X_{t} \tanh \Gamma_{t}\, dt \, + \, 2
\sqrt{\omega} \sigma_{q}\,
d W_{t}, \qquad\quad & \tilde{X}_{0} = 0 \; \text{m},\\
d \tilde{K}_{t} & = & 2 \lambda X_{t} \tanh \Gamma_{t}\, dt \, + \,
2 \sqrt{\lambda}\, d W_{t},  & \tilde{K}_{0} = 0 \; \text{m}^{-1},
\end{array}
\end{equation}
where $X_{t}$ is given by Eq.~\eqref{eq:n2}. This is a non-linear
system, since it depends in a non-linear way on $\Gamma_{t}$, which
is also a stochastic process; as such, (to our knowledge) the system
can not be exactly solved. To circumvent this problem, let us
consider the following two auxiliary linear systems:
\begin{equation} \label{eq:linsys}
\begin{array}{lcll}
d \tilde{X}^{\pm}_{t} & = & \displaystyle
\frac{\hbar}{m}\,\tilde{K}^{\pm}_{t}\, dt \, \pm\, \omega X_{t}\, dt
\, + \, 2 \sqrt{\omega} \sigma_{q}\,
d W_{t}, \qquad\quad & \tilde{X}^{\pm}_{0} = 0 \; \text{m},\\
d \tilde{K}^{\pm}_{t} & = & \pm 2 \lambda X_{t}\, dt \, + \, 2
\sqrt{\lambda}\, d W_{t},  & \tilde{K}^{\pm}_{0} = 0 \;
\text{m}^{-1}.
\end{array}
\end{equation}
(with an obvious meaning of the symbols), which have been obtained
in the first case ($+$) by replacing $\tanh \Gamma_{t}$ with $+1$,
and in the second case ($-$) by replacing $\tanh \Gamma_{t}$ with
$-1$. Clearly, we have: $\tilde{X}^{-}_{t} \leq \tilde{X}_{t} \leq
\tilde{X}^{+}_{t}$ and $\tilde{K}^{-}_{t} \leq \tilde{K}_{t} \leq
\tilde{K}^{+}_{t}$ for any $t$ such that $X_{t} \geq 0$, which is
true for all the time during which the experiment takes place, and
much longer. Such linear systems can be easily solved; concerning
$\tilde{X}^{\pm}_{t}$, and after some tedious calculations one finds
the following time dependence for the mean:
\begin{equation} \label{eq:fddfgd}
{\mathbb E}_{\mathbb P} [ \tilde{X}^{\pm}_{t} ] \; = \; \pm \,
\left\{
\begin{array}{ll}
- X_{t} + \hbar \kappa t \qquad & \text{for $t < T$}, \\
- X_{t} + \hbar \kappa T \qquad & \text{for $t \geq T$},
\end{array}
\right.
\end{equation}
and for the variance:
\begin{equation} \label{eq:fvar}
{\mathbb V}_{\mathbb P} [ \tilde{X}^{\pm}_{t} ] \; = \;
4\sigma_{q}^{2} \left[ \omega t + \frac{(\omega t)^2}{2} +
\frac{(\omega t)^3}{12} \right].
\end{equation}
We use the above results to approximate the time evolution of
$\tilde{X}_{t}$ and thus of $\bar{x}^{+}_{t}$ and $\bar{x}^{-}_{t}$,
which we are interested in. We consider separately the case $t \leq
T_{C}$ (before the collapse) and $t \geq T_{C}$ after the collapse:
in the first case, we cannot control the behavior of $\Gamma_{t}$,
thus the most we can say is that $|\tanh\Gamma_{t}| \leq 1$, which
has already been used to bound $\tilde{X}_{t}$ between
$\tilde{X}^{-}_{t}$ and $\tilde{X}^{+}_{t}$; in the second case, we
know that with very high probability $|\tanh\Gamma_{t}| \geq \tanh
35$, which is a very strong bound.

\noindent {\it Case 1, before the collapse: $t \leq T_{C}$.} Within
this time interval, the two Gaussian components $|G +, t\rangle$ and
$|G -, t\rangle$ start separating, as $X_{t}$ increases in time; in
particular, at time $t = T_{C}$, when the collapse has (almost
certainly) occurred, we have:
\begin{equation} \label{eq:dfdfg1}
{\mathbb E}_{\mathbb P} [ \tilde{X}^{\pm}_{T_{C}} ] \quad \simeq
\quad \pm \,\frac{1}{2}\,\hbar \kappa \omega T_{C}^{2} \quad \simeq
\quad \pm 5.9 \times 10^{-15} \text{m},
\end{equation}
which has being obtained from Eq.~\eqref{eq:fddfgd} by expanding
$X_{t}$, as given by Eq.~\eqref{eq:n2}, to second order in $\omega
t$; moreover, we have from Eq.~\eqref{eq:fvar}:
\begin{equation}
{\mathbb V}_{\mathbb P} [ \tilde{X}^{\pm}_{T_{C}} ] \quad \simeq
\quad 4 \omega \sigma_{q}^2 T_{C} \quad \simeq \quad 6.5 \times
10^{-35} \text{m}^{2}.
\end{equation}
This means that, on a macroscopic scale, $\tilde{X}^{\pm}_{T_{C}}
\simeq {\mathbb E}_{\mathbb P} [ \tilde{X}^{\pm}_{T_{C}} ]$; since
$X_{T_{C}} \simeq \hbar \kappa T_{C} \simeq 1.5 \times 10^{-6}$ m
$\gg {\mathbb E}_{\mathbb P} [ \tilde{X}^{\pm}_{T_{C}} ]$, and
keeping in mind that $\tilde{X}^{-}_{T_{C}} \leq \tilde{X}^{\phantom
\pm}_{T_{C}} \leq \tilde{X}^{+}_{T_{C}}$, we can write, with very
high probability and very good approximation:
\begin{eqnarray}
\bar{x}^{+}_{T_{C}} & \simeq & +\frac{1}{2}\, X_{T_{C}} \; \simeq \;
+\frac{1}{2}\, \hbar \kappa T_{C} \; \simeq \; +7.7 \times 10^{-7}
\;
\text{m}, \\
\bar{x}^{-}_{T_{C}} & \simeq & -\frac{1}{2}\, X_{T_{C}} \; \simeq \;
-\frac{1}{2}\, \hbar \kappa T_{C} \; \simeq \; -7.7 \times 10^{-7}
\; \text{m}.
\end{eqnarray}
Accordingly, and as expected, the two components move symmetrically
in opposite directions, one towards the right and the other towards
the left, but not fast enough for a macroscopic superposition to
occur, before the collapse enters into play and suppresses one of
them.

\noindent {\it Case 2, after the collapse: $t \geq T_{C}$.} Let us
assume that the collapse is such that the outcome ``$+\hbar/2$'' is
given; this means that almost certainly $\Gamma_{t} \geq 35$,
$\forall \,\, t \geq T_{C}$. Given this, let us first of all show
that $\tilde{X}_{t}$ remains very close to $\tilde{X}^{+}_{t}$, for
very long times; then, by approximating $\tilde{X}_{t}$ with
$\tilde{X}^{+}_{t}$, we will show how $\bar{x}^{+}_{t}$ and
$\bar{x}^{-}_{t}$ evolve in time.

From Eqs.~\eqref{eq:nlinsys} and~\eqref{eq:linsys}, taking into
account that $\tanh \Gamma_{t} \geq -1$, we find:
\begin{eqnarray}
\tilde{K}^{+}_{t} - \tilde{K}^{\phantom \pm}_{t} & = & 2 \lambda
\int_{0}^{t} X_{t'} (1 - \tanh \Gamma_{t'}) \, dt' \; \leq \; 4
\lambda \int_{0}^{t} X_{t'} \, dt' \; = \nonumber \\
& = & \frac{8\lambda \hbar \kappa}{\omega^2} \left[ 1 - e^{-\omega
t/2} \left(\cos \frac{\omega t}{2} + \sin \frac{\omega t}{2} \right)
\right] \; \simeq \; 2\lambda \hbar \kappa t^{2}
\end{eqnarray}
and
\begin{eqnarray}
\tilde{X}^{+}_{t} - \tilde{X}^{\phantom \pm}_{t} & = &
\frac{\hbar}{m} \int_{0}^{t} (\tilde{K}^{+}_{t'} -
\tilde{K}^{\phantom \pm}_{t'})\, dt' + \omega \int_{0}^{t} X_{t'} (1
- \tanh \Gamma_{t'}) \,
dt' \nonumber \\
& \leq & \frac{\hbar}{m} \int_{0}^{t} (\tilde{K}^{+}_{t'} -
\tilde{K}^{\phantom \pm}_{t'})\, dt' + 2 \omega \int_{0}^{t} X_{t'}
\, dt' \; \leq \; -2(X_{t} - \hbar \kappa t) \simeq \hbar \kappa
\omega t^{2}.
\end{eqnarray}
At time $t = T_{C}$, we then have: $\tilde{K}^{+}_{T_{C}} -
\tilde{K}^{\phantom \pm}_{T_{C}} \simeq 2\lambda \hbar \kappa
T_{C}^{2} \simeq 2.8 \times 10^{12}$ m$^{-1}$, and:
$\tilde{X}^{+}_{T_{C}} - \tilde{X}^{\phantom \pm}_{T_{C}} \simeq
\hbar \kappa \omega T_{C}^{2} \simeq 1.2 \times 10^{-14}$ m.

We use these results as initial conditions, at time $T_{C}$, to
find, by integrating once more Eqs.~\eqref{eq:nlinsys}
and~\eqref{eq:linsys}, and by using the two inequalities $\tanh
\Gamma_{t} \geq \eta \equiv \tanh 35$, $\forall\,\, t \geq T_{C}$
and $X_{t} \leq \ell \simeq 1$ cm, the following estimates:
\begin{eqnarray}
\tilde{K}^{+}_{t} - \tilde{K}^{\phantom \pm}_{t} & = &
\tilde{K}^{+}_{T_{C}} - \tilde{K}_{T_{C}} \, + \, 2 \lambda
\int_{T_{C}}^{t} X_{t'} (1 - \tanh \Gamma_{t'}) \, dt' \; \leq \;
\tilde{K}^{+}_{T_{C}} - \tilde{K}_{T_{C}} \, + \, 2 \lambda \eta
\ell (t - T_{C})
\end{eqnarray}
and
\begin{eqnarray}
\tilde{X}^{+}_{t} - \tilde{X}^{\phantom \pm}_{t} & \leq &
\tilde{X}^{+}_{T_{C}} - \tilde{X}^{\phantom \pm}_{T_{C}}  \, + \,
\frac{\hbar}{m} \int_{T_{C}}^{t} (\tilde{K}^{+}_{t'} -
\tilde{K}^{\phantom \pm}_{t'})\, dt' + \omega \int_{T_{C}}^{t}
X_{t'} (1 - \tanh \Gamma_{t'}) \, dt' \nonumber \\
& \leq & \tilde{X}^{+}_{T_{C}} - \tilde{X}^{\phantom \pm}_{T_{C}} +
\frac{\hbar}{m}(\tilde{K}^{+}_{T_{C}} - \tilde{K}^{\phantom
\pm}_{T_{C}})(t - T_{C}) + \frac{\omega^2}{4} \eta \ell (t -
T_{C})^2 \, + \,
\omega \eta \ell (t - T_{C})  \nonumber \\
& \simeq & \omega\hbar\kappa T_{C}^{2} + \frac{\omega^2}{2}\, \hbar
\kappa T_{C}^{2} (t - T_{C}) + \frac{\omega^2}{4}\, \eta \ell (t -
T_{C})^2 + \omega \eta
\ell (t - T_{C}) \nonumber \\
& \simeq & \left( 1.2 \times 10^{-14} + 2.9 \times 10^{-19}t + 5.0
\times 10^{-42} t^2 \right) \text{m},
\end{eqnarray}
We see that for very long times, by far much longer than the time
during which the experiment takes place, the distance between
$\tilde{X}^{\phantom \pm}_{T_{C}}$ and $\tilde{X}^{+}_{T_{C}}$
remains small, so small that we can replace $\tilde{X}^{\phantom
\pm}_{T_{C}}$ with $\tilde{X}^{+}_{T_{C}}$ for all practical
purposes.

On the other hand, $\tilde{X}^{+}_{T_{C}}$ is, on a macroscopic
scale, very close to its average value ${\mathbb E}_{\mathbb P} [
\tilde{X}^{+}_{t} ]$, its variance, as given by Eq.~\eqref{eq:fvar},
being extremely small; accordingly we have:
\begin{equation}
\bar{x}^{+}_{t} \; = \; \frac{X_{t} + \tilde{X}_{t}}{2} \; \simeq \;
\frac{X_{t} + \tilde{X}^{+}_{t}}{2} \; \simeq \; \frac{X_{t} +
{\mathbb E}_{\mathbb P} [ \tilde{X}^{+}_{t} ]}{2} \; = \; \left\{
\begin{array}{ll}
\displaystyle +\frac{\hbar \kappa t}{2} \qquad & t < T, \\
& \\ \displaystyle +\frac{\hbar \kappa T}{2} & t \geq T,
\end{array}
\right.
\end{equation}
which is the desired result: the pointer, represented in this case
by $|G +, t\rangle$, moves at a constant speed towards the right and
stops at the position $\hbar \kappa T/2$, displaying in this way the
correct outcome.

To conclude the analysis, let us see what happens also to the other
component, $|G -, t\rangle$, which has been suppressed by the
spontaneous reduction process. Its center $\bar{x}^{-}_{t}$ moves
approximately as follows:
\begin{equation}
\bar{x}^{-}_{t} \; = \; -\frac{X_{t} - \tilde{X}_{t}}{2} \; \simeq
\; -\frac{X_{t} - \tilde{X}^{+}_{t}}{2} \; \simeq \; -\frac{X_{t} -
{\mathbb E}_{\mathbb P} [ \tilde{X}^{+}_{t} ]}{2} \; \simeq \;
\left\{
\begin{array}{rl}
\displaystyle - \frac{\hbar \kappa t}{2} & \qquad t < T, \\
& \\ - \displaystyle \frac{\hbar \kappa T}{2} & \qquad T \leq t \ll
\omega^{-1}, \\ &
\\
 \displaystyle  + \frac{\hbar \kappa T}{2} & \qquad t
 \gg \omega^{-1}.
\end{array}
\right.
\end{equation}
i.e. the negligible Gaussian component moves to the left of the
graduate scale, but then slowly converges towards the other wave
function.

As a final remark, we note that, at very long times of order
$\omega^{-1} \simeq 2.0 \times 10^{4}$ sec, the statistical
fluctuations become relevant also on the macroscopic scale, thus
approximating any actual value with its statistical average becomes
less and less precise. However, times of order $\omega^{-1} \simeq
2.0 \times 10^{4}$ sec are by far much longer that the time required
for the experiment to end; moreover, for such long times the
assumption that the global system is isolated certainly looses its
validity; the measurement model should then be refined, in order to
include so long time scales.

\section{Conclusions}
\label{sec:conclusions}

In the present work we have analyzed the quantum theory of
measurement within the framework of dynamical reduction models,
resorting to the von~Neumann type scheme of measurement process and
to the QMUPL model of spontaneous wave function collapse. We have
proven the properties listed in the introductory section, showing in
this way how the axioms 4 and 5 of standard Quantum Mechanics arise
in quite a straightforward way from the dynamical evolution law
governing models of spontaneous wave function collapse.

We hope that our analysis makes clearer the mechanism with which
dynamical reduction models provide, at least at the non relativistic
level, such an accurate description of measurement processes, and
more generally of all physical situations.

\begin{acknowledgments}
We acknowledge very stimulating discussion with D. D\"urr and
G.C.~Ghirardi. The work of A.B. has been partly supported by the EU
grant MEIF CT 2003--500543 and partly by DFG (Germany). D.G.M.S.
gratefully acknowledges support from Istituto Nazionale di Fisica
Nucleare, Sezione di Trieste, and the Department of Theoretical
Physics of the University of Trieste.
\end{acknowledgments}

\appendix
\section{A mistake corrected in Ref.~\cite{ab1}}

When $\kappa = 0$, i.e. for a free particle of mass $m$ moving
according to the SDE:
\begin{equation} \label{nle}
d\,\psi_{t}(x) \; = \; \left[ -\frac{i}{\hbar}\, \frac{p^2}{2m}\, dt
+ \sqrt{\lambda}\, (q - \langle q \rangle_{t})\, dW_{t} -
\frac{\lambda}{2}\, (q - \langle q \rangle_{t})^2 dt \right]
\psi_{t}(x),
\end{equation}
the two equations~\eqref{eq:19} and~\eqref{eq:20} for $\gamma_{t}$
and $\theta_{t}$, respectively, become (we neglect the $\pm$):
\begin{eqnarray}
\label{eq:ap1} d\gamma_{t} & = & \left( \lambda \bar{x}_{t}^{2} +
\frac{\hbar}{m}\, \alpha^{\makebox{\tiny I}}_{t} + \frac{\lambda}{4
\alpha^{\makebox{\tiny R}}_{t}}\right)dt \; + \;  \sqrt{\lambda}\,
\bar{x}_{t} \left\{ d\xi_{t} - 2\sqrt{\lambda}\, \bar{x}_{t}\, dt
\right\}\\
\label{eq:ap2} d\theta_{t} & = & \left( - \frac{\hbar}{2m}\,
\bar{k}_{t}^{2} - \frac{\hbar}{m}\, \alpha^{\makebox{\tiny R}}_{t} +
\frac{\lambda \alpha^{\makebox{\tiny I}}_{t}}{4
(\alpha^{\makebox{\tiny R}}_{t})^2} \right)dt +  \sqrt{\lambda}
\frac{\alpha^{\makebox{\tiny I}}_{t}}{\alpha^{\makebox{\tiny
R}}_{t}}\,\bar{x}_{t} \left\{ d\xi_{t} - 2\sqrt{\lambda}\,
\bar{x}_{t}\, dt \right\}, \;\;\;\;\;
\end{eqnarray}
which differ from the corresponding Eqs.~(12) and~(13)
of~\cite{ab1}, in the first case for the extra factor $\lambda/4
\alpha^{\makebox{\tiny R}}_{t}$ and in the second case for the
factor $\lambda \alpha^{\makebox{\tiny I}}_{t}/4
(\alpha^{\makebox{\tiny R}}_{t})^2$. We correct in this way a
mistake made in Ref.~\cite{ab1}, which however does not affect the
other results contained in that paper.

\end{document}